\documentstyle[12pt,a4,epsf,rotate]{article}

\newlength{\notewidth}
\newcommand{\lt}{\left}
\newcommand{\rt}{\right}
\newcommand{\ov}{\overline}
\newcommand{\nn}{\nonumber \\}
\newcommand{\no}{\nonumber }
\newcommand{\T}{{\, \rm \bf T \,}}
\newcommand{\g}{\gamma}
\newcommand{\ns}{\mbox{no sum on }}
\newcommand{\lef}{\lt( 1-\g_5 \rt) }

\newcommand{\ba}{\mbox{\tiny bare} }
\newcommand{\re}{\mbox{\tiny ren} }
\newcommand{\tr}{\mbox{tr} }
\newcommand{\da}{\frac{\partial }{\partial a_{il}}}
\newcommand{\e}{\varepsilon}

\newcommand{\Lagr}{{\cal L}}
\newcommand{\bare}{\ba}
\newcommand{\ren}{\re}
\newcommand{\diff}{{\rm d}}
\newcommand{\eq}[1]{(\ref{#1})}
\newcommand{\eps}{\varepsilon}
\newcommand{\imag}{{\rm Im}\,}

\begin{document}
% SH's footnote style during the title
\renewcommand{\thefootnote}{\fnsymbol{footnote}}

{\sf
\begin{minipage}{2in}
\begin{flushleft}
	TUM-T31-66/94 \\
	PSI-PR-94-37
\end{flushleft}
\end{minipage}
\hfill
\begin{minipage}{2in}
\begin{flushright}
	hep-ph/9412375 \\
	August 1995
\end{flushright}
\end{minipage}
\vspace{3ex} \\
}
\begin{center}
	{\LARGE
	Evanescent Operators, Scheme Dependences and Double Insertions
	\footnote{Work supported by the German
		``Bundesministerium f\"ur Bildung, Wissenschaft,
                  Forschung und Technologie''
		under contract no.\ 06-TM-743.}
	}
\end{center}
\vspace{1pc}
\begin{center}
	{\large Stefan Herrlich
	\footnote{e-mail: {\tt herrl@feynman.t30.physik.tu-muenchen.de}}
	} \\
	{\sl Paul Scherrer Institut, CH-5232 Villigen PSI, Switzerland}
	\\[2pc]
	{\large Ulrich Nierste
	\footnote{e-mail: {\tt nierste@feynman.t30.physik.tu-muenchen.de}}
	} \\
	{\sl Physik-Department, TU M\"unchen, D-85747 Garching, Germany}
\end{center}
\vspace{2pc}

% SH's style settings...
\setcounter{footnote}{0}
\renewcommand{\thefootnote}{\arabic{footnote})}

\begin{abstract}
The anomalous dimension matrix of dimensionally regularized four-quark
operators is known to be affected by evanescent operators, which
vanish in $D=4$ dimensions.
Their definition, however, is not unique, as one can always redefine
them by adding a term proportional to $(D-4)$ times a physical operator.
In the present paper we compare different definitions used in the
literature and find that they correspond to different renormalization
schemes in the {\em physical} operator basis.
The scheme transformation formulae for the Wilson coefficients and the
anomalous dimension matrix are derived in the next-to-leading order.
We further investigate the proper treatment of evanescent operators in
processes appearing at second order in the effective four-fermion
interaction such as particle-antiparticle mixing, rare hadron decays
or inclusive decays.
\end{abstract}

\section{Introduction}
\label{Sect:Intro}
In the past two decades much effort has been made to calculate QCD
corrections to weak processes.
The indispensable renormalization group improvement of perturbatively
calculated Feynman amplitudes requires their factorization into
Wilson coefficients and matrix elements, which are obtained from an
effective field theory containing four-fermion interactions.
When calculating QCD radiative corrections to these four-fermion
operators using dimensional regularization ($D=4-2 \varepsilon $) one
faces evanescent Dirac structures such as
\begin{eqnarray}
\g_{\mu} \g_{\nu} \g_{\vartheta} (1-\g_5) \otimes
\g^{\vartheta} \g^{\nu} \g^{\mu } (1-\g_5) -
(4+a \varepsilon)  \g_{\mu} (1-\g_5) \otimes \g^{\mu} (1-\g_5),
\label{introex}
\end{eqnarray}
which vanish in $D=4$ dimensions, but  appear with a factor of
$1/\varepsilon$ in counterterms to physical operators.
By introducing the parameter $a$ in \eq{introex} we have displayed the
arbitrariness in the definition of the evanescent operators:  A priori
one can add any multiple of $\varepsilon $ times any physical operator
to a given evanescent operator. In the literature one indeed finds
different definitions of the latter.
The consequences of this arbitrariness for renormalization group
improved Green's functions are one of the main subjects of this paper.

The role of evanescent operators in perturbation theory has  been
investigated since the pioneering era of dimensional regularization
\cite{b,c,bm}.
When perturbative results are to be improved by means of the operator
product expansion and  renormalization group (RG) techniques to sum
large logarithms, new subtleties arise:
First the matrix elements of evanescent operators can affect the
matching equation determining the Wilson coefficients which multiply
the effective four-fermion operators \cite{bw}.
Second the appearance of evanescent operators in counterterms to
physical operators and vice versa leads to the mixing of physical and
evanescent operators during the RG evolution \cite{c,dg}.
In \cite{c,bw} a finite renormalization of the evanescent operators
has been proposed to render their matrix elements zero.
By this the Wilson coefficients of the evanescent operators become
irrelevant at the matching scale. For this to be true at any scale it
is important that simultaneously the evanescent operators do not mix
into physical ones.
In \cite{dg} it has been proven for a very special definition of the
bare evanescent operators that this is indeed the case, if the finite
renormalization proposed in \cite{c,bw} is performed.
But does this feature hold for any definition of the evanescent
operators, i.e.\ for any choice of $a$ in \eq{introex}?
We will affirm this question in section~\ref{Sect:Triang} after
setting up our notations and describing and generalizing the commonly
used definitions of evanescent operators in section~\ref{Sect:Prelim}.

It is well-known that a change in the renormalization prescription of
the composite operators affects the Wilson coefficients and the
anomalous dimension matrix in the next-to-leading order (NLO) and
beyond.
In section~\ref{Sect:Scheme} we will find that a change in the
definition of the evanescent operators, i.e.\ a change of $a$ in
\eq{introex}, leads to a different form of the {\em physical} part of
the anomalous dimension matrix. Hence a different $a$ corresponds to a
different renormalization scheme.
This result is of utmost practical importance for any calculation
beyond leading logarithms, as it shows that it is meaningless to state
a result for some anomalous dimension matrix without mentioning the
definition of the evanescent operators used in the calculation.
If one wants to combine some anomalous dimension matrix with Wilson
coefficients or perturbative matrix elements calculated with a
different definition of the evanescent operators, one clearly needs
scheme transformation formulae for the Wilson coefficients and the
anomalous dimension matrix.
We will derive these scheme transformation formulae in the
next-to-leading order in section~\ref{Sect:Scheme}, too.

When studying particle-antiparticle mixing or rare decays one faces
Green's functions with two insertions of four-fermion operators.
The second main subject of this paper is to work out the correct
treatment of these Green's functions when one or both inserted
operators are evanescent.
In section~\ref{Sect:Double} we extend the results of \cite{bw,dg} and
of sections~\ref{Sect:Triang} and \ref{Sect:Scheme} to this case of
double insertions.

Then in section~\ref{Sect:Inclusive} inclusive decays are discussed.
We close our paper with our conclusions.

\section{Preliminaries and Notation}
\label{Sect:Prelim}
Let $\{ \hat{Q}_k=\ov{\psi} q_k \psi \cdot \ov{\psi} \tilde{q}_k \psi,
\; k=1,2,3,\ldots \}$ be a set of physical dimension-six
four-quark operators. We are interested in the  Green's functions
of a SU(N) gauge theory with insertions of $\hat{Q}_k$
renormalized by minimal subtraction (${\rm MS}$).
The arguments are easily generalized to other mass-independent
renormalization schemes like $\overline{\rm MS}$.
The Dirac structures $Q_k = q_k \otimes \tilde{q}_k$ corresponding to
$\hat{Q}_{k}$ are considered to form a basis of the space of Lorentz
singlets and pseudosinglets for $D=4$.
Neither the Lorentz indices of $q_k$ and $\tilde{q}_k$ are
displayed nor any flavour or colour indices, which are irrelevant for the
discussion of the subject.
$\lt[ \Gamma \otimes 1 \rt] Q_k \lt[ 1 \otimes \Gamma ^\prime \rt]$
means
$\Gamma q_k \otimes \tilde{q}_k \Gamma ^\prime$.
Frequently we will use the example
\begin{eqnarray}
Q&=&\g_{\mu } \lef \otimes \g^{\mu } \lef.
\label{q}
\end{eqnarray}

The matrix elements of  $\hat{Q}_k$ have some perturbative expansion
in the gauge coupling $g$:
\begin{eqnarray}
Z_\psi ^2
\langle \hat{Q}_k^{\ba}  \rangle
  &=&
 \sum _{j \geq 0} \lt( \frac{g^2}{16 \pi ^2}  \rt)^j
\langle \hat{Q}_k ^{\ba} \rangle ^{(j) }. \label{me}
\end{eqnarray}
Here $Z_{\psi}$ is the quark wave function renormalization constant.
The right hand side of (\ref{me}) still contains divergences, which
are to be removed by the renormalization of the operators
$\hat{Q}_{k}$ \cite{zim}.

\begin{figure}[htb]
\centerline{
\rotate[r]{\epsfysize=12cm \epsffile{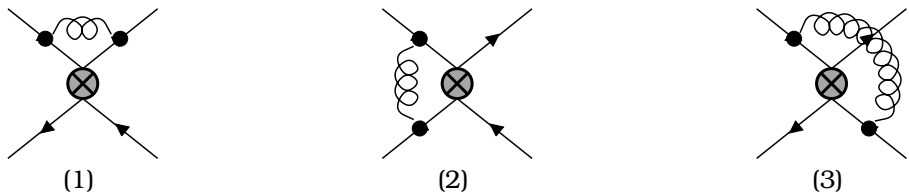 }}
} \caption{Diagrams contributing to
$\protect\langle \protect\hat{Q}_k ^{\protect\ba} \protect\rangle
^{(1)}$.}
\label{Fig:1}
\end{figure}
Now the insertion of $\hat{Q}_k$ into the one-loop diagrams of
fig.~\ref{Fig:1} yields a linear combination of the $\hat{Q}_l$'s and
a new operator with the Dirac structure
$Q_k^\prime= \lt[ \g_\rho \g_\sigma  \otimes 1 \rt]
  Q_k \lt[ 1 \otimes \g^\sigma \g^\rho  \rt]$:
\begin{eqnarray}
\langle \hat{Q}_k ^{\ba} \rangle ^{(1) } &=&
d^{(1)}_{kl} \langle \hat{Q}_l \rangle ^{(0)} +
d^{(1)}_{k , Q^\prime _k}
   \langle \hat{Q}^\prime _k \rangle ^{(0)} \quad \ns k \label{d1},
\end{eqnarray}
where $\langle \ldots \rangle^{(0)}$ denote tree level matrix elements.
Both coefficients have a term proportional to $1/\varepsilon $ and a finite
part. $\hat{Q}_k^\prime$ is now decomposed into a linear combination of
the $\hat{Q}_l$'s and an evanescent operator:
\begin{eqnarray}
\hat{Q}_k^{\prime \ba} &=& \lt( f_{kl} + a_{kl} \varepsilon \rt)
   \hat{Q}_l ^{\ba }
      + \hat{E}_1[ Q_k]^{\ba } \label{e1}
+O(\varepsilon ^2).
\label{DefEvan1}
\end{eqnarray}
Here the $f_{kl}$'s  are uniquely determined by the
Dirac basis decomposition
in $D=4$ dimensions. The
$a_{kl}$'s, however, are arbitrary, and a different choice for the
$a_{kl}$'s  corresponds to a different definition of
$\hat{E}_1 [ Q_k ]=\hat{E}_1 [ Q_k, \{ a_{rs} \} ]$.
%The part of the gauge boson propagator involving the gauge parameter
%does not contribute to $d^{(1)}_{k , Q^\prime _k} $ in (\ref{d1}) and
%is therefore irrelevant for the discussion of evanescent operators.
%Hence we can adopt the Landau gauge in the following, so that
%the wave function renormalization equals unity.
When going beyond the one-loop order new evanescent operators
$\hat{E}_2 [ Q_k ], \hat{E}_3 [ Q_k ], \ldots $
will appear. Their precise definition is irrelevant for the moment and
will be given after (\ref{defe2}).

Now in the framework of dimensional regularization the renormalization
of some physical operator $\hat{Q}_k$ requires counterterms
proportional to physical and evanescent operators:
We define the renormalization matrix $Z$  by
\begin{eqnarray}
Z_\psi ^2 \langle \hat{Q}_k ^{\ba} \rangle   &=& Z_{kl} \langle
      \hat{Q}_l ^{\re} \rangle  +
  Z_{k,E_{jm}} \langle \hat{E}_j [Q_m] ^{\re} \rangle  \quad.
\label{z}
\end{eqnarray}
Here and in the following we will distinguish the renormalization
constants related to some evanescent operator $E_j[Q_m]$ by denoting
the corresponding index with $E_{jm}$.
(\ref{d1}) and (\ref{e1}) imply that $Z_{kl}^{(1)}$ depends on the
$a_{rs}$'s,
while $Z_{k,E_{jm}}^{(1)}$ is independent of
them. We define  the coefficients in the expansion of
$Z$ in terms of the gauge coupling constant $g$ and in
terms of $\varepsilon $ by
\begin{eqnarray}
Z &=& 1+ \sum _{j} \lt( \frac{g^2 }{16 \pi ^2} \rt)^j  Z ^{(j)} ,
\quad \quad
Z^{(j)} \, = \,
        \sum _{k=0}^{j} \frac{1}{\varepsilon ^k} Z ^{(j)} _{k}
\label{exp}.
\end{eqnarray}

The first analysis of evanescent operators in the context of RG
improved QCD corrections to electroweak transitions has been done
by Buras and Weisz \cite{bw}.
They have determined the  $a_{kl}$'s by choosing some
set of Dirac structures $M= \{ \g^{(1)} \otimes \tilde{\g} ^{(1)}, \ldots
 \g^{(10)} \otimes \tilde{\g} ^{(10)}  \} $,
which forms a basis for $D=4$, and contracting all elements in $M$ with
$Q_k^\prime$ and $Q_l$ in (\ref{e1}):
\begin{eqnarray}
\tr \left( \g^{(m)} q_k ^\prime \tilde{\g} ^{(m)} \tilde{q}_k ^\prime
\right)  &=&
(f_{kl} + a_{kl} \varepsilon ) \,
\tr \left( \g^{(m)} q_l \tilde{\g}^{(m)} \tilde{q}_l  \right)
+O(\varepsilon ^2) , \no \\
&& \quad \quad \quad
 \ns k \mbox{ and on $m$\/=1,\ldots , 10}.
\label{gp}
\end{eqnarray}
The solution of the  equations (\ref{gp}) uniquely defines the
$f_{kl} + a_{kl} \varepsilon $.
In other words, $E_1[Q_k ]$ obeys the equations:
\begin{eqnarray}
E_1[Q_k]_{ijrs} \g^{(m)}_{si} \tilde{\g}^{(m)}_{jr} &=&
 O ( \varepsilon^2 ) \quad \quad \quad
\mbox{ for $m$\/=1, \ldots ,10},
\end{eqnarray}
where $i,j,r,s$ are Dirac indices.

Our arguments will not depend on the scheme used for the treatment of
$\g_5$. In the examples we will use a totally anticommuting $\g_5$.
This does not cause any ambiguity in the trace operation in
(\ref{gp}), because all Lorentz indices are contracted, so that the
traced Dirac string is a linear combination of $\g_5$
and the unit matrix.

E.g.\ the choice of
\begin{eqnarray}
M&=& \{ 1\otimes1, 1\otimes \g_5, \g_5 \otimes 1, \g_5 \otimes \g _5 ,
   \g_{\mu} \otimes \g ^{\mu},
   \g_{\mu} \otimes \g ^{\mu} \g_5, \nn
&& \quad
   \g_5 \g_{\mu} \otimes \g^{\mu},
   \g_5 \g_{\mu} \otimes \g^{\mu} \g_5,
   \sigma_{\mu \nu } \otimes \sigma ^{\mu \nu },
   \g_5 \sigma_{\mu \nu } \otimes \sigma ^{\mu \nu }
         \}  \label{m}
\end{eqnarray}
gives  for $Q$ in (\ref{q})
\begin{eqnarray}
Q^\prime &\! = \!  &
 \g_\rho \g_\sigma
\g_{\mu } \lef \otimes \g^{\mu } \g^\sigma   \g^\rho \lef  =
( 4- 8 \varepsilon ) Q + E_1[Q]
 +O(\varepsilon^2) \label{ex}
\end{eqnarray}
as in \cite{bw}. We remark that this choice $a=-8$ respects the Fierz
symmetry, which relates the first to the second diagram in
fig.~\ref{Fig:1}.

A basis different from $M$ in (\ref{m}) yields the same
$f_{k,l}$'s, but different $a_{kl}$'s. For example
by replacing the sixth and eighth element of $M$ in (\ref{m}) by
$\g_\alpha \g_\beta \g_\delta \otimes \g ^\alpha \g^\beta \g^\delta $
and
$\g_5 \g_\alpha \g_\beta \g_\delta \otimes \g ^\alpha \g^\beta \g^\delta $
one finds
\begin{eqnarray}
Q^\prime &=&  4  Q + 16 \varepsilon
   (1+\g_5) \otimes (1-\g_5) + E_1^\prime [Q_k] +O(\varepsilon^2), \no
\end{eqnarray}
instead of (\ref{ex}),
i.e.\   a different evanescent operator.
The Dirac algebra is infinite dimensional for non-integer $D$ and is
spanned by $M$ and an infinite set of evanescent Dirac structures.
Hence one can reverse the above procedure  and first arbitrarily choose the
$a_{kl}$'s and then add properly  adjusted
 linear combinations of the evanescent
structures  to the elements of $M$ such as to obtain the chosen $a_{kl}$'s.

Yet the so defined evanescent operators do not decouple from the
physics in four dimensions:
In \cite{c,bw} it  has    been observed that their one-loop
 matrix elements generally have nonvanishing components proportional
to the physical operators $Q_k$:
\begin{eqnarray}
\langle \hat{E}_1 [Q_k ] ^{\ba} \rangle ^{(1)}  &=&
     \lt[ Z^{(1)}_0 \rt]_{E_{1k},l}  \langle \hat{Q}_l \rangle ^{(0)} +
       \frac{1}{\varepsilon }
     \lt[ Z^{(1)}_1 \rt]_{E_{1k},E_{1k} }
   \langle \hat{E}_1 [Q_k ] \rangle ^{(0)} \nn
&&   +  \frac{1}{\varepsilon }
     \lt[ Z^{(1)}_1 \rt]_{E_{1k},E_{2k} }
   \langle \hat{E}_2 [Q_k ] \rangle ^{(0)}
  +O(\varepsilon) \label{e2}.
\end{eqnarray}
Here  a second evanescent operator $\hat{E}_2$, which will be discussed in a
moment,  has appeared.
Clearly no sum on $k$ is
understood in (\ref{e2}) and in following analogous places.
In (\ref{e2}) $[Z^{(1)}_0]_{E_{1k},l}$ is
local, because it originates from the local  $1/\varepsilon$--pole
of the tensor integrals and a term proportional to $\varepsilon$
stemming from the evanescent Dirac algebra.
For the same reason there is no divergence in the term proportional to
$\langle \hat{Q}_l \rangle ^{(0)} $.
Now in \cite{c,bw} it has been proposed to renormalize $\hat{E}_{1}$ by a
finite amount to cancel this component:
\begin{eqnarray}
  \hat{E}_{1} [Q_k ]  ^{ \re }    &=&
  \hat{E}_{1} [Q_k ]  ^{ \ba }  + \frac{g^2}{16 \pi ^2 } \lt\{
     - \lt[ Z^{(1)}_0 \rt]_{E_{1k},l}   \hat{Q}_l  \rt.
\nn
&& \quad \quad
-     \frac{1}{\varepsilon }
     \lt[ Z^{(1)}_1 \rt]_{E_{1k},E_{1k} }
     \hat{E}_1 [Q_k]      \nn
&& \quad \quad \lt.  -    \frac{1}{\varepsilon }
     \lt[ Z^{(1)}_1 \rt]_{E_{1k},E_{2k} }
    \hat{E}_2 [Q_k ]  \rt\} +O \lt( g^4 \rt).
\label{fin}
\end{eqnarray}
With (\ref{fin}) the renormalized matrix elements
of the evanescent operators are $O(\varepsilon )$, so that they
do not contribute to the one-loop matching of some  Green's function
$G^{\re}$ in the full renormalizable theory
with matrix elements in the effective theory:
\begin{eqnarray}
 i G^{\re } &=& C_l \langle \hat{Q}_{l}  \rangle ^{\re}
     + C_{ E_{1k} } \langle \hat{E}_{1} [Q_k]  \rangle ^{\re}
     + O \lt( g^4  \rt), \label{match}
\end{eqnarray}
i.e.\ the coefficients $C_{ E_{1k} }$ are irrelevant, because they
multiply  matrix elements which vanish for $D=4$.
In \cite{bw} it has been further noticed
that $ Z ^ {(1)}_0 $ in
(\ref{fin}) influences the two-loop anomalous dimension matrix of the
{\em physical} operators, so that the presence of evanescent operators
indeed has an  impact on physical observables.
In a different context this has also been observed in \cite{bos}.

Next we discuss $\hat{E}_2[Q_k]$, which has entered the scene in
(\ref{e2}): When inserting
$\hat{E}_1[Q_k]$ defined in (\ref{e1})
into the one-loop diagrams of fig.~\ref{Fig:1}, one involves
\begin{eqnarray}
Q_k^{\prime \prime} &=&
\lt[ \g_\rho \g_\sigma  \otimes 1 \rt]
  Q_k^\prime \lt[ 1 \otimes \g^\sigma \g^\rho  \rt]
\nn
&=&  \lt[ f +a \varepsilon  \rt]^2_{\, kl}
                       Q_l
      + \lt( f_{kl} + a_{kl} \varepsilon \rt) E_1[Q_l] \nn
&&  +  \lt[ \g_\rho \g_\sigma  \otimes 1 \rt]
  E_1 [ Q_k ] \lt[ 1 \otimes \g^\sigma \g^\rho  \rt]
 \label{ins} \\
&=&      \lt\{ \lt[ f + a \varepsilon \rt]^2 _{kl} + b_{kl}
              \varepsilon \rt\} Q_l
  + E_2 [Q_k]
  + O\lt( \varepsilon ^2   \rt)
    , \label{defe2}
\end{eqnarray}
which defines $E_2[Q_k]= E_2[Q_k,\{ a_{rs} \} , \{ b_{rs} \}   ]$.
Only the last term in
(\ref{ins}) can contribute to the
new coefficients $b_{kl}$. If the projection is performed
with e.g.\ $M$ defined in (\ref{m}), one finds
$b_{kl}=0$
\footnote{This is the case for any basis $M$ in which for each
	$\g^{(m)} \otimes \g^{(m)} \in M$ the quantity
	$\g_\rho \g_\sigma \g^{(m)} \g^\sigma \g^\rho \otimes \g^{(m)}$
	is a linear combination of the elements in $M$.}.
In our discussion we will keep $b_{kl}$ arbitrary.
Clearly, one has a priori to deal with the mixing of an infinite
set of evanescent operators $\lt\{ \hat{E}_j[Q_k] \rt\}$ for each physical
operator $\hat{Q}_k$,
where $\hat{E}_{j+1}[Q_k]$ denotes the new evanescent operator
appearing first in the one-loop matrix elements of $\hat{E}_j[Q_k]$.

With the finite renormalization of $\hat{E}_1[Q_k]$ in
(\ref{fin}) the evanescent operators  do not affect the
physics at the matching scale,
at which   (\ref{match}) holds.
In order that this will be true at any scale $\mu $, however,
one must
also ensure that the evanescent operators do not mix into the physical ones.
 This has been noticed first by Dugan and Grinstein in \cite{dg}.
For the operator basis
$(\vec{Q},\vec{E})
	=(\hat{Q}_1,\ldots \hat{Q}_n , \,
	  \hat{E}_1 \lt[ Q_1 \rt], \ldots \hat{E}_j \lt[ Q_k \rt], \ldots )$
this means that the anomalous dimension matrix
\begin{eqnarray}
\g &=& \lt(
\begin{array}{cc}
\g_{QQ} &  \g_{QE} \\
\g_{EQ} &  \g_{EE}
\end{array} \rt)
\nonumber
\end{eqnarray}
has an upper block-triangular form with $\g_{EQ}=0$.

The authors of \cite{dg} have introduced another way to define the
evanescent operators, which is also frequently used:
It is easy to see that one can restrict the operator basis $\{Q_k\}$
to the set of operators whose Dirac structures $q_k, \tilde{q}_k$
are completely antisymmetric in their Lorentz indices.
This is the normal form of Dirac strings introduced in \cite{bm}.
Dirac strings being antisymmetric in more than four indices vanish in
four dimensions and are therefore evanescent.
Operators with five antisymmetrized indices correspond to $\hat{E}_1$
in our notation, and $\hat{E}_2$ would be expressed in terms of a
linear combination of Dirac structures with seven and with five
antisymmetrized indices.
Clearly this method also corresponds to some special choice for the
$a_{kl}$'s and $b_{kl}$'s in (\ref{e1}) and (\ref{defe2}).
Now in \cite{dg}
the authors have proven that with the use of those definitions and a
finite renormalization analogous to (\ref{fin}) the anomalous
dimension matrix indeed has the desired block-triangular form, so that
the evanescent operators do not mix into the physical ones.
While the anomalous dimension matrix is trivially block-triangular
at one-loop level, the proof for the  two-loop level was given in
\cite{dg} by the use of the abovementioned special definition of the
evanescent operators.
The latter, however, has some very special features, which are absent
for the general case with arbitrary  $a_{kl}$'s and $b_{kl}$'s, e.g.\
the definition used in  \cite{dg} automatically yields an anomalous
dimension matrix which is tridiagonal in the evanescent sector.

Consider now a definition of the evanescent operators different from
the one used in \cite{dg}: By inserting the definition (\ref{e1}) of
$\hat{E}_1 \lt[ Q_k \rt]$ into (\ref{e2}) one realizes that $\lt[
Z^{(1)}_0 \rt]_{E_{1k},l}$ depends on the $a_{kl}$'s. Similarly at the
two-loop level $\lt[ Z_1^{(2)} \rt]_{E_{jk},l} $ depends on the
definition (\ref{e1}), so that one has to wonder which choices for the
$a_{kl}$'s lead to the desired block-triangular form of $\g$ with
$\g_{EQ}=0$.  In the following section we will prove that any choice
is permissible. Further we will find that also the $b_{kl}$'s may be
chosen completely arbitrary.

On the other hand the physical submatrix $\g_{QQ}$ of the anomalous
dimension matrix depends on the $a_{kl}$'s as we will show in
section~\ref{Sect:Scheme}.  Hence the freedom in the definition of the
bare evanescent operators induces a renormalization scheme dependence
in the physical sector of the operator basis. This feature has not
been discussed in the literature so far.  As emphasized in the
introduction it is of practical importance for NLO calculations to
know the scheme transformation formulae for the physical submatrix
$\g_{QQ}$ and the Wilson coefficients. We will come back to this point
in section~\ref{Sect:Scheme}.

\section{Block Triangular Anomalous Dimension Matrix}
\label{Sect:Triang}
Consider some set of physical operators $\{ \hat{Q}_k \}$ which closes
under renormalization together with the corresponding evanescent
operators $\{ \hat{E}_j[Q_k] : j \geq 1 \}$. Their $O(\varepsilon)$--parts
$a_{rs},b_{rs}, \ldots $ are chosen arbitrarily.  We want to show that
the block of the anomalous dimension matrix describing the mixing of
$\hat{E}_j[Q_k]$ into $\hat{Q}_l$ equals zero,
\begin{eqnarray}
\lt[ \g \rt]_{E_{jk},l} &=& 0,   \label{nomix}
\end{eqnarray}
provided one uses the finite renormalization described in (\ref{fin}).

Our sketch will follow the outline of \cite{dg}, where
(\ref{nomix}) has been proven by complete induction.
At the one-loop level (\ref{nomix}) is trivial, and the induction starts
in two-loop order:
The next-to-leading order contribution to the anomalous dimension
matrix
$\gamma = \frac{g^2}{16\pi^2} \gamma^{\left(0\right)}
	+ \left(\frac{g^2}{16\pi^2}\right)^2 \gamma^{\left(1\right)}
	+ \ldots$
reads \cite{bw}:
\begin{eqnarray}
\g^{(1)} &=& -4     Z_1^{(2)}  -2 \beta_0 Z_0^{(1)} +
2 \lt\{  Z_1^{(1)} , Z_0^{(1)}      \rt\}. \label{twomatrix}
\end{eqnarray}
The nonzero  contributions to (\ref{nomix}) in two-loop order are
\begin{eqnarray}
	\lt[ \g^{(1)} \rt]_{ E_{jk},l }
	&=&
	-4  \lt[ Z_1^{(2)} \rt]_{ E_{jk},l }
	-2 \beta_0 \lt[ Z_0^{(1)} \rt] _{ E_{jk},l }
\nn
	& &
	+ 2 \lt\{
		\lt[ Z_1^{(1)} \rt]_{ E_{jk},E_{rs} }
		\lt[ Z_0^{(1)}\rt] _{ E_{rs},l }
		+
		\lt[ Z_0^{(1)}  \rt] _{ E_{jk},m }
		\lt[ Z_1^{(1)}  \rt] _{ m l  }
	    \rt\}.
\label{twoloop}
\end{eqnarray}
Here (\ref{twoloop}) contains terms which are absent when the  special
definition of the evanescent operators in \cite{dg} is used:
In \cite{dg} one has $\lt[ Z^{(1)} \rt]_{E_{jk},l}=0 $
for $j \geq 2$ contrary to  the general case, where any evanescent
operator can have counterterms proportional to physical operators.

Next we look at $\lt[ Z_1^{(2)} \rt]_{E_{jk},l}$, which stems from the
$1/\varepsilon$--term of the $O(g^4)$--matrix elements
of $\hat{E}_j[Q_k]$. As discussed in \cite{dg}, these $1/\varepsilon$--terms
originate from
$1/\varepsilon^2$--poles in the tensor integrals
multiplying a factor proportional to $\varepsilon$ stemming from the
evanescent Dirac algebra.
Now in each two-loop diagram the former are related to the
corresponding one-loop counterterm diagrams by a factor of 1/2,
because the non-local $1/\varepsilon$--poles cancel in their sum
\cite{ho}.
For this to hold it is crucial that the one-loop counterterms are
properly adjusted, i.e.\ that they cancel the
$1/\varepsilon$--poles in the one-loop tensor integrals.
In the one-loop matrix elements of evanescent operators the latter
are multiplied with $\varepsilon$ originating from the Dirac algebra.
Hence the proper one-loop renormalization of the evanescent operators
must be such as to give matrix elements of order $\e$, as shown
for $E_1[Q_k]$ in (\ref{fin}).

{From} the  one-loop counterterm graphs one simply reads off:
\begin{eqnarray}
\lt[ Z_1^{(2)}  \rt]_{E_{jk},l}
\!\!\!  &=& \!\!
   \frac{1}{2}  \lt\{
      \lt[ Z_0^{(1)}  \rt]_{E_{jk},m}  \!
      \lt[ Z_1^{(1)}  \rt]_{ ml} \! \!
+  \lt[ Z_1^{(1)}  \rt]_{E_{jk},E_{rs}} \!
\lt[ Z_0^{(1)}  \rt]_{E_{rs},l} \! \!
- \beta_0
\lt[ Z_0^{(1)}  \rt]_{E_{jk},l}  \rt\},    \no
\end{eqnarray}
which yields the desired result when inserted into (\ref{twoloop}).
Here the first two terms stem from insertions of physical and
evanescent counterterms to $E_j[Q_k]$, while the term involving
the coefficient of the one-loop $\beta$--function
$\beta (g) = - g^3/(16 \pi^2) \beta_0 $
originates from the diagrams with coupling constant counterterms.
The terms involving the  wave function renormalization constants
cancel with those stemming from the factor  $Z_\psi^2$ in
(\ref{z}).

The inductive step in \cite{dg} proving
(\ref{nomix}) to any loop order does not use any special definition
of the evanescent and therefore applies unchanged here.

\section{Evanescent Scheme Dependences}
\label{Sect:Scheme}
In this section we will analyze  the dependence of the
physical part of $\g^{(1)}$ given in (\ref{twomatrix})
and of the one-loop Wilson coefficients on
$a_{il}$ and $b_{il}$.
In practical next-to-leading order calculations one often has to
combine Wilson coefficients and anomalous dimension matrices
obtained with different definitions of the evanescent operators
and it is therefore important to have formulae allowing to
switch between them (see e.g.\ appendix B of \cite{m}).

We start with the investigation of the dependence of
$\g ^{(1)}$ on $a_{il}$.
The bare one-loop matrix element
\begin{eqnarray}
\langle \hat{Q}_k ^{\ba} \rangle^{(1) }  &=&
\lt\{
  \frac{1}{\varepsilon}  \lt[ Z_1^{(1)}  \rt]_{kj}
  +  \lt[ d_0^{(1)} \rt]_{kj} \rt\}  \langle  \hat{Q}_j \rangle^{(0)}
+ \frac{1}{\varepsilon} \lt[ Z_1^{(1)}  \rt]_{k,E_{1k}}
  \langle  \hat{E}_1 [Q_k]   \rangle ^{(0)} \nn
&&+ O( \varepsilon)
\label{q1}
\end{eqnarray}
is independent of $a_{il}$, which is evident from
(\ref{d1}).
$E_1[Q_k]$ depends linearly on $a_{il}$ through its definition
(\ref{e1}) with the coefficient
\begin{eqnarray}
\da \hat{E}_1[Q_k] &=& - \varepsilon \, \delta _{ki} \,  \hat{Q}_l,
\label{DiffE1}
\end{eqnarray}
so that  (\ref{q1}) gives:
\begin{eqnarray}
\da \lt[ d_0^{(1)} \rt]_{kj}  &=& \lt[ Z_1^{(1)} \rt]_{k,E_{1k}}
\delta _{ki} \delta_{lj}, \label{d0}
\end{eqnarray}
while $Z_1^{(1)}$ is independent of $a_{il}$.

In the same way on can obtain the $a_{kl}$--dependence
of $Z_1^{(2)}$. (\ref{z}) reads to two-loop order (cf.~(\ref{me})):
\begin{eqnarray}
\langle \hat{Q}_k ^{\ba} \rangle^{(2) }
&=& Z_{kj}^{(2)}
    \langle \hat{Q}_j  \rangle^{(0)}
+   Z_{k,E_{1m}}^{(2)}  \langle \hat{E}_1 [Q_m]   \rangle^{(0) }
+   Z_{k,E_{2m}}^{(2)}  \langle \hat{E}_2 [Q_m]   \rangle^{(0) }
\nn
&&   + Z_{kr}^{(1)}  \langle \hat{Q}_r ^{\re} \rangle^{(1) }
+  Z_{k,E_{1k}}^{(1)} \langle \hat{E}_1 [Q_k] ^{\re}  \rangle^{(1)}
+   \langle \hat{Q}_k ^{\re} \rangle^{(2) }.
\label{q2}
\end{eqnarray}
{From} (\ref{defe2}) we know
\begin{eqnarray}
\da \hat{E}_2 [Q_m] &=& - \varepsilon \lt[ f_{mi} \delta _{lj} +
\delta_{mi} f_{lj}    \rt] \hat{Q}_j ,
\end{eqnarray}
and from (\ref{q1}) one reads off:
\begin{eqnarray}
    \langle \hat{Q}_r ^{\re} \rangle^{(1) }  &=&
 \lt[ d_0^{(1)} \rt]_{rj}     \langle \hat{Q}_j  \rangle^{(0)}  \label{qren}.
\end{eqnarray}
These relations and (\ref{d0}) allow to calculate the derivative
of  (\ref{q2}) with respect to $a_{il}$.
Keeping in mind that the evanescent matrix elements are $O(\e )$
the $O(1/ \e)$--part of the derivative yields:
\begin{eqnarray}
\da \lt[ Z_1^{(2)}  \rt] _{kj}  &=&
-  \lt[ Z_1^{(1)}  \rt]_{ki}
  \lt[ Z_1^{(1)}  \rt]_{i,E_{1i}} \delta_{lj} +
  \lt[ Z_2^{(2)}  \rt]_{k,E_{1i}}  \delta_{lj} \nn
&& + \lt[ Z_2 ^ {(2)}  \rt]_{k,E_{2m}}
 \lt( \delta_{mi} f_{lj} + f_{mi} \delta_{lj}    \rt)
,\quad \ns i. \label{z2}
\end{eqnarray}

Again $\lt[ Z_2^{(2)} \rt]$
can be extracted from the one-loop counterterm diagrams as described
in the preceding section:
\begin{eqnarray}
\lt[ Z_2^{(2)}  \rt]_{k,E_{1i}}
&=&  \frac{1}{2}
\lt[ Z_1^{(1)}  \rt]_{ki} \lt[ Z_1^{(1)}  \rt]_{i,E_{1i} } +
 \frac{1}{2}
\lt[ Z_1^{(1)}  \rt]_{i,E_{1i}} \lt[ Z_1^{(1)}  \rt]_{E_{1i},E_{1i} }
\delta _{ki} \nn
&& - \, \frac{1}{2} \beta _0
\lt[ Z_1^{(1)}  \rt]_{i,E_{1i} } \delta_{ki} , \quad \quad \ns i  \nn
\lt[ Z_2^{(2)}  \rt]_{k,E_{2m}}
&=&
 \frac{1}{2}
\lt[ Z_1^{(1)}  \rt]_{k,E_{1k}} \lt[ Z_1^{(1)}  \rt]_{ E_{1k},E_{2k} }
\delta_{km}
 , \quad \quad \ns k.
\label{coun}
\end{eqnarray}
After inserting (\ref{coun}) into (\ref{z2}) we want to
substitute the last term in (\ref{z2}).  For this we
derive both sides of (\ref{e2})
with respect to $a_{il}$
giving:
\begin{eqnarray}
\lefteqn{ \lt[ Z_1^{(1)}   \rt]_{E_{1k} ,E_{2k} }
\lt( \delta_{ki} f_{lj} + f_{ki} \delta_{lj}    \rt)
\; =} \nn
&& \da \lt[ Z_0^{(1)}   \rt]_{E_{1k},j}
 +  \lt[ Z_1^{(1)}   \rt]_{lj} \delta_{ki}  -
\lt[ Z_1^{(1)}   \rt]_{E_{1k} ,E_{1k} } \delta_{ki} \delta_{lj}, \label{z0}
\quad  \ns k.
\end{eqnarray}
Finally one has to insert the expression for  (\ref{z2})
obtained by the described substitutions
into
\begin{eqnarray}
\da \lt[ \g^ {(1)} \rt]_{kj} &=&
-4 \da \lt[ Z_1^{(2)}   \rt]_{kj}
+ 2 \lt[ Z_ 1 ^ {(1)}  \rt]_{k,E_{1k}}
\da \lt[ Z_0^{(1)} \rt] _{E_{1k},j}, \quad \ns k , \no
\end{eqnarray}
which follows from (\ref{twomatrix}).
The result reads:
\begin{eqnarray}
\da \lt[ \g ^{(1)} \rt]_{kj} &=&
- 2 \lt[ Z_1^{(1)} \rt]_{lj} \lt[ Z_1^{(1)} \rt]_{i,E_{1i}} \delta _{ki}
 + 2 \lt[ Z_1^{(1)} \rt]_{ki} \lt[ Z_1^{(1)} \rt]_{i,E_{1i}}
  \delta_{lj} \nn
&& +
 2 \beta_0 \lt[ Z_1^{(1)} \rt]_{i,E_{1i}} \delta_{ki} \delta_{lj},
      \quad \quad \ns i.  \label{dag}
\end{eqnarray}
Since the quantities on the right hand side of (\ref{dag}) do not
depend on $a_{il}$, one can easily integrate (\ref{dag}) to find
the desired relation between two $\g$'s corresponding to  different
% definitions of the evanescent operators corresponding
choices for $a_{kl}$
in (\ref{e1}).
To write the result in matrix form we recall the expression for the
physical one-loop anomalous dimension matrix
\begin{eqnarray}
\lt[ \g ^{(0)} \rt]_{lj} &=& -2 \lt[ Z_1^{(1)} \rt] _{lj} \no
\end{eqnarray}
and introduce the diagonal matrix $D$ with
\begin{eqnarray}
D_{im} &=&  \lt[ Z_1^{(1)} \rt] _{i,E_{1i}} \delta_{im}
   , \quad \quad \quad \ns i.
\label{DiagMatrix1}
\end{eqnarray}
Hence
\begin{eqnarray}
\g ^{(1)} ( a^\prime)&=&
\g ^{(1)} (a ) + \lt[ D \cdot ( a^\prime -a ), \g^{(0)} \rt]
       + 2 \, \beta_0 \, D \cdot (a^\prime -a)  ,   \label{result}
\end{eqnarray}
where the summation in the row and column indices only runs over
the physical submatrices.

(\ref{result}) exhibits  the familiar structure of the scheme
dependence of $\g ^{(1)}$ \cite{bjlw}. Usually scheme dependences
are analyzed for a fixed definition of the bare operators and
different subtraction procedures.
%, and (\ref{result}) can  easily be derived from the definition of $\g$.
Our situation, however, is more
complicated, because we investigate the scheme dependence associated
with different definitions of the {\em bare} operator basis
(i.e.\ of the bare evanescent operators).

The dependence of the one-loop matrix elements on $a$ can be found
easily from (\ref{qren}) and (\ref{d0}):
\begin{eqnarray}
\langle \vec{Q} \rangle ^{\re}  (a^\prime) &=&
 \lt[ 1+ \frac{g^2}{16 \pi^2} D \cdot (a^\prime-a)    \rt]
  \langle \vec{Q}  \rangle ^{\re} (a) + O( g^4).
\label{depma}
\end{eqnarray}
Since in (\ref{match}) $G$ does not depend on $a$ and the evanescent
matrix element is $O(\e )$, the corresponding relation for the
Wilson coefficients at the matching scale reads:
\begin{eqnarray}
 \vec{C}^T  (a^\prime) &=&
   \vec{C}^T  (a)
 \lt[ 1 -  \frac{g^2}{16 \pi^2} D \cdot (a^\prime-a)    \rt]
+ O( g^4).
\label{SchemeWC1}
\end{eqnarray}
Hence we can apply the result of \cite{bjlw}, which shows
that in the renormalization group improved Wilson coefficient
the scheme dependences in (\ref{result})  cancels the one in
 (\ref{depma}),
so that physical observables are scheme independent,
provided the hadronic matrix elements are defined scheme
independently.

Let us drop some words on the results \eq{result},
\eq{depma} and \eq{SchemeWC1}:
In general one would expect scheme transformation formulae involving
the full operator basis $(\vec{Q},\vec{E})$.  Yet all summations only
run over the indices corresponding to the physical operators, the only
ingredient from the evanescent sector being the matrix $D$.  This is
why we could not simply deduce \eq{result} from \eq{depma} using the
results of \cite{bjlw}.  Possible contributions from summations over
evanescent operator indices in the matrix products in \eq{result}
cannot be inferred from
\eq{depma}, because there they would multiply vanishing matrix
elements.

In the same way one can investigate the dependence of
$\g^{(1)}$ on $b_{il}$ given in (\ref{defe2}): While $Z_1^{(2)}$
and $Z_0^{(1)}$ depend on $b_{il}$, this dependence cancels in
(\ref{twomatrix}). Hence  neither $\g^{(1)}$ nor the one-loop
Wilson coefficient are affected by the choice of $b_{il}$.

In general $\g^{(0)}$ and $\g^{(1)}$ do not commute, so that
one has to cope with
complicated matrix equations in order to solve
the renormalization group equation
in next-to-leading order \cite{bjlw}.
Now one can use (\ref{result}) to  simplify $\g^{(1)}$: By going to
the diagonal basis for
\mbox{$\g^{(0)}=\mbox{diag}\left(\g^{(0)}_i\right)$}
one can easily find solutions for $a^\prime - a$ in (\ref{result})
which even give \mbox{$\g^{(1)}(a^\prime)=0$} provided that all
$Z_{k,E_{1k}}$'s are nonzero and all eigenvalues of $\g^{(0)}$
satisfy \mbox{$\left| \g^{(0)}_i -\g^{(0)}_j \right| \neq 2 \beta_0$}.
We will exemplify this in a moment.

A choice for $a_{il}$ which leads to a $\g^{(1)}$ commuting with
$\g^{(0)}$ has been done implicitly  in \cite{bw}:
There the mixing of the two operators
$Q_+ = Q \lt( {\rm 1} + \tilde{\rm 1} \rt) /2 $ and
$Q_- = Q \lt( {\rm 1} - \tilde{\rm 1} \rt) /2 $ has been considered,
where ${\rm 1}$ and $\tilde{\rm 1}$ denote colour singlet
and antisinglet and $Q$ was introduced in (\ref{q}).
Now $Q_+$ is self-conjugate under the Fierz transformation,
while $Q_-$ is anti-self-conjugate, so that
$\g^{(0)}$ is diagonal to maintain the
Fierz symmetry in the leading order renormalization group
evolution.
As remarked after (\ref{ex}), the definition of $E_1[Q]$
in (\ref{ex}) is necessary to ensure the  Fierz symmetry
in the  one-loop matrix elements.  Consequently
with (\ref{ex}) also $\g^{(1)}$ has to obey the Fierz symmetry
preventing the mixing of $Q_+$ and $Q_-$, i.e.\ yielding
a diagonal $\g^{(1)}$. A different definition of $E_1[Q]$
would  result in non-Fierz-symmetric matrix elements,
but in renormalization scheme independent expressions
they would combine with  a non-diagonal $\g^{(1)}$
such as to restore Fierz symmetry.

Let us consider the example above to demonstrate that one can
pick $a^\prime$ such that $\gamma ^{(1)} (a ^\prime ) =0$:
In  \cite{bw} the definitions
\begin{eqnarray}
 E_1[ Q_\pm] & =&
 \lt(   \pm
1- \frac{1}{N} \rt)
   \lt[  \g_\rho \g_\sigma \g_\nu (1-\g_5) \otimes
     \g^\nu \g^\sigma \g^\rho (1-\g_5) \rt.  \nn
&& \quad \quad \lt.   -  (4 - 8 \varepsilon)\;\;
  \g_\nu (1-\g_5) \otimes \g ^\nu (1-\g_5)
    \rt], \no
\end{eqnarray}
i.e.\ $a_{++}=8 (1/N-1), a_{--}=8 (1/N+1), a_{+-}=a_{-+}=0 $,  were
adopted yielding a diagonal
$\g^{(1)}(a)=$diag$\lt(\g^{(1)}_{+}(a),\g^{(1)}_{-}(a)\rt)$.
{From} the insertion of $Q_+$ and $Q_-$ into the diagrams of
fig.~\ref{Fig:1} one finds $Z_{+,E_{1+}}=Z_{-,E_{1-}}=1/4$.
Hence if we pick
$a^\prime_{\pm \pm}=a_{\pm \pm} - 2/\g^{(1)}_{\pm} (a)/\beta_0$ and
$a^\prime_{\pm \mp}=0$, (\ref{result}) will imply
$\g^{(1)}(a^\prime)=0$.

% here starts SH's work
\section{Double Insertions}
\label{Sect:Double}

% first, some redefinitions...
\renewcommand{\nn}{\no}

\subsection{Motivation}
\label{Sect:DoubleMotiv}

In the following we will investigate Green functions with two
insertions of local operators.
Consider first the effective Lagrangian to first order written in
terms of bare local operators
\begin{eqnarray}
	\Lagr^{\rm I}
	&=&
	C_{k} Z^{-1}_{kl} \hat{Q}^{\bare}_{l}
	+
	C_{k} Z^{-1}_{k E_{rl}} \hat{E}_{r}\left[Q_{l}\right]^{\bare}
\nn \\
	& &
	+
	C_{E_{jk}} Z^{-1}_{E_{jk} l} \hat{Q}^{\bare}_{l}
	+
	C_{E_{jk}} Z^{-1}_{E_{jk} E_{rl}} \hat{E}_{r}\left[Q_{l}\right]^{\bare}
	.
\label{Lagr1}
\end{eqnarray}
According to the procedure presented in the preceding sections, the
coefficients $C_{E_{jk}}$ were found to be irrelevant and therefore
remained undetermined.

\begin{figure}[htb]
	\centerline{
	\rotate[r]{
		\epsfysize=5cm
		\epsffile{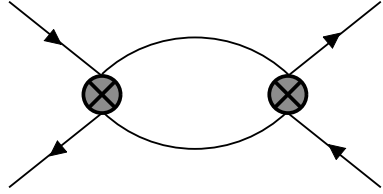}
	}}
	\caption{The lowest order diagram contributing to the Green
	function in \protect\eq{GFdouble}.}
\label{Fig:2}
\end{figure}

Now consider 4-fermion Green functions with insertion of two local
operators from $\Lagr^{\rm I}$
\begin{eqnarray}
	\Bigl\langle 0 \Bigr|
	\T
	\bar{\Psi}_{1}
	\bar{\Psi}_{2}
	\left(
	\frac{i}{2} \int \diff^{D}y
	\Lagr^{\rm I}\left(x\right)
	\Lagr^{\rm I}\left(y\right)
	\right)
	\Psi_{3}
	\Psi_{4}
	\Bigl| 0 \Bigr\rangle
	.
\label{GFdouble}
\end{eqnarray}
Such Green functions appear in applications like particle-antiparticle
mixing or rare hadron decays.
The diagram contributing to lowest order is depicted in fig.~\ref{Fig:2}.
Renormalization of them in general requires additional counterterms
proportional to new local dimension-eight operators $\tilde{Q}_{l}$,
because the diagram of fig.~\ref{Fig:2} is divergent:
\begin{eqnarray}
	\Lagr
	&=&
	\Lagr^{\rm I}
	+
	\Lagr^{\rm II}
\label{Lagr}
\\
	\Lagr^{\rm II}
	&=&
	C_{k} C_{k'}
	\left\{
		Z^{-1}_{kk',l} \tilde{Q}^{\bare}_{l}
		+
		Z^{-1}_{kk',E_{rl}}
			\tilde{E}_{r}\left[\tilde{Q}_{l}\right]^{\bare}
	\right\}
\nn \\
	& &
	+
	C_{k} C_{E_{j' k'}}
	\left\{
		Z^{-1}_{k E_{j' k'},l} \tilde{Q}^{\bare}_{l}
		+
		Z^{-1}_{k E_{j' k'},E_{rl}}
			\tilde{E}_{r}\left[\tilde{Q}_{l}\right]^{\bare}
	\right\}
\nn \\
	& &
	+
	C_{E_{j k}} C_{E_{j' k'}}
	\left\{
		Z^{-1}_{E_{jk} E_{j' k'},l} \tilde{Q}^{\bare}_{l}
		+
		Z^{-1}_{E_{jk} E_{j' k'},E_{rl}}
			\tilde{E}_{r}\left[\tilde{Q}_{l}\right]^{\bare}
	\right\}
\nn \\
	& &
	+
	\tilde{C}_{k} \tilde{Z}^{-1}_{kl} \tilde{Q}^{\bare}_{l}
	+
	\tilde{C}_{k} \tilde{Z}^{-1}_{k E_{rl}}
		\tilde{E}_{r}\left[\tilde{Q}_{l}\right]^{\bare}
\nn \\
	& &
	+
	\tilde{C}_{E_{jk}} \tilde{Z}^{-1}_{E_{jk} l} \tilde{Q}^{\bare}_{l}
	+
	\tilde{C}_{E_{jk}} \tilde{Z}^{-1}_{E_{jk} E_{rl}}
		\tilde{E}_{r}\left[\tilde{Q}_{l}\right]^{\bare}
\label{Lagr2}
\end{eqnarray}
Here $Z^{-1}_{..,.} \tilde{Q}_{.} $ are the local operator
counterterms needed to renormalize the divergences originating purely
from the double insertion.
Further we have explicitly distinguished physical and evanescent
operators.
The renormalization constants $Z_{..,.}$, clearly being symmetric in
their first two indices, give rise to an inhomogeneity in the RG
equation for the Wilson coefficients $\tilde{C}_{k}$,
$\tilde{C}_{E_{rs}}$, which we call the anomalous dimension tensor of
the double insertion.
Note, that this quantity also has three indices, see \eq{AnomDimTens}.
It has become standard to define the local operator $\tilde{Q}_{l}$
with inverse powers of the coupling constant such that
$Z^{-1}_{..,.}=O\left(g^2\right)$ to avoid mixing already at the tree
level.
As an example take
$\tilde{Q}=\frac{m^2}{g^2}\;
	\gamma_{\mu} \left(1-\gamma_5\right) \otimes \gamma^{\mu}
	\left(1-\gamma_5\right)$
for which $\tilde{C}_{l}=O\left(g^2\right)$.
For simplicity, we assume the $\tilde{Q}_{l}$'s to be linearly
independent from the $\hat{Q}_{k}$'s
\footnote{e.g.\ the $\hat{Q}_{k}$'s represent $\Delta F = 1$
	operators, the $\tilde{Q}_{l}$'s denote $\Delta F = 2$
	operators, where $F$ is some quantum number, which is
	conserved by the SU(N) interaction.}.
The $E_{r}\left[\tilde{Q}_{l}\right]$ in \eq{Lagr2} are defined
analogously to \eq{DefEvan1} with new coefficients $\tilde{f}_{kl}$,
$\tilde{a}_{kl}$, $\tilde{b}_{kl}$, etc.
Hence new arbitrary constants $\tilde{a}_{kl}$, $\tilde{b}_{kl}$
potentially causing scheme dependences enter the scene.

Clearly the following questions arise here:
\begin{enumerate}
\item
\label{Q1}
Are the coefficient functions $C_{E_{jk}}$ irrelevant also for the
double insertions;
i.e.\ do
\begin{eqnarray}
\left\langle \int \T \hat{E} \hat{Q} \right\rangle
\hspace{1cm} \mbox{and} \hspace{1cm}
\left\langle \int \T \hat{E} \hat{E} \right\rangle
\label{DoubleWithEva}
\end{eqnarray}
contribute to the matching procedure and the operator mixing?
\item
\label{Q2}
Does one need a {\em finite} renormalization in the evanescent sector
of double insertions;
if yes, how does this affect the anomalous dimension tensor?
\item
\label{Q3}
How do the $\tilde{C}_{l}$ and anomalous dimension matrices depend on
the $a_{kl}$, $b_{kl}$, $\tilde{a}_{kl}$, $\tilde{b}_{kl}$ ?
\item
\label{Q4}
Are the RG improved observables scheme independent?
\end{enumerate}

\subsection{Scheme Consistency}
\label{Sect:DoubleConsist}
In this section we will carry out the program of
section~\ref{Sect:Triang} for the case of double insertions to answer
questions~\ref{Q1} and \ref{Q2} (on page~\pageref{Q1}).

Two cases have to be distinguished:
The matrix element of the double insertion of the two local
renormalized operators can be divergent or finite:
\begin{eqnarray}
	\left\langle \frac{i}{2} \int \T \hat{Q}_{k} \hat{Q}_{k'}
		\right\rangle
	&=&
	\left\{
	\begin{array}{lcl}
	\mbox{divergent} & , & \mbox{case 1} \\
	\mbox{finite} & , & \mbox{case 2}
	\end{array}
	\right.
	.
\label{cases}
\end{eqnarray}
Case~1 is the generic one, appearing in the calculation of the
coefficient $\eta_{3}$ in ${\rm K}^{0}$--$\overline{{\rm K}^{0}}$
mixing \cite{hn2} or in ${\rm K} \to \pi \nu \bar{\nu}$ \cite{bb}.
Case~2 appears, if the divergent parts of different contributions to
\eq{cases} add such that the divergences cancel.
It is realized e.g.\ in the determination of $\eta_{1}$ in
${\rm K}^{0}$--$\overline{{\rm K}^{0}}$ mixing \cite{hn}.
Therefore we need or do not need a separate renormalization for the
double insertion
\begin{eqnarray}
	Z^{-1}_{k k',l}
	\left\{
	\begin{array}{clcl}
	\neq & 0 & , & \mbox{case 1} \\
	=    & 0 & , & \mbox{case 2}
	\end{array}
	\right.
	.
\end{eqnarray}
Since we need an extra renormalization in case~1, let us introduce the
symbol
$\left[ \frac{i}{2} \int \T Q Q \right]^{\ren}$
for the completely renormalized operator product constructed from two
renormalized local operators $Q$ with an additional renormalization
factor for the double insertion.

Let us start the discussion with the matching procedure:
At some renormalization scale we have to match Green functions
obtained in the full theory with Green functions calculated in the
effective theory:
\begin{eqnarray}
	-i G^{\ren}
	&=&
	C_{k} C_{k'} \left\langle \left[\frac{i}{2} \int \T
		\hat{Q}_{k} \hat{Q}_{k'} \right]^{\ren} \right\rangle
	+
	C_{k} C_{E_{i' k'}} \left\langle \left[i \int \T
		\hat{Q}_{k} \hat{E}_{i'}\left[Q_{k'}\right]
		\right]^{\ren} \right\rangle
\nn \\
	& &
	+
	C_{E_{ik}} C_{E_{i' k'}} \left\langle \left[\frac{i}{2} \int
		\T \hat{E}_{i}\left[Q_{k}\right]
		\hat{E}_{i'}\left[Q_{k'}\right]
		\right]^{\ren} \right\rangle
	+
	C_{l} \left\langle \tilde{Q}_{l} \right\rangle
\nn \\
	& &
	+
	\tilde{C}_{E_{jl}} \left\langle
		\tilde{E}_{j}\left[\tilde{Q}_{l}\right]
		\right\rangle
	,
\end{eqnarray}
where $G^{\ren}$ corresponds e.g.\ to a ``box'' function in the full
SM.
Since the coefficients $C_{E_{jk}}$ must be irrelevant for this
matching procedure, one must have
\begin{eqnarray}
	Z_{\psi}^{2}
	\left\langle \left[ \frac{i}{2} \int \T
	\hat{E}_{j}\left[Q_{k}\right] \hat{Q}_{k'} \right]^{\ren}
	\right\rangle
	&\stackrel{!}{=}&
	\left\{
		\begin{array}{l@{\hspace{1cm}}l}
		O\left(\eps^{0}\right) & \mbox{in case 1 (LO)} \\
		O\left(\eps^{1}\right) & \mbox{in case 1 (NLO and higher)} \\
		O\left(\eps^{1}\right) & \mbox{in case 2}
		\end{array}
	\right.
\nn \\
	& &
\label{CondMatchDouble}
\end{eqnarray}
and analogously for two insertions of evanescent operators.
To understand this recall that the purpose of RG improved perturbation
theory is to sum logarithms. In case~1 the LO matching is performed by
the comparison of the coefficients of logarithms of the full theory
amplitude and the effective theory matrix element \eq{CondMatchDouble}
(the latter being trivially related to the coefficient of the
divergence),
while the NLO matching is obtained from the finite part and also
involves the matrix elements of the local operators \cite{hn2,bb}.
In case~2 the matching is performed with the finite parts in all
orders \cite{hn}.
In both cases the condition \eq{CondMatchDouble} is trivially
fulfilled in LO, because the evanescent Dirac algebra gives an
additional $\eps$ compared to the case of the insertion of two
physical operators.
Therefore a finite renormalization for the double insertion turns out
to be unnecessary at the LO level.
This statement remains valid at the NLO level only in case~2, in
case~1 condition \eq{CondMatchDouble} no longer holds if one only
subtracts the divergent terms in the matrix elements containing a
double insertion.
With the argumentation preceding \eq{fin} one finds that in this case
the finite term needed to satisfy the condition \eq{CondMatchDouble}
is local and therefore can be provided by a finite counterterm.

The operator mixing is more complicated.
To deal with this, we need the evolution equation for the Wilson
coefficient functions $\tilde{C}_{k}$, $\tilde{C}_{E_{rs}}$, which
can be easily derived from the renormalization group invariance of
$\Lagr^{\rm II}$ and reads
\begin{eqnarray}
	\mu \frac{\diff}{\diff \mu} \tilde{C}_{l}
	&=&
	\tilde{C}_{l'} \tilde{\gamma}_{l' l}
	+
	C_{k} C_{n} \gamma_{k n,l}
\label{inh}
\end{eqnarray}
with the anomalous dimension tensor of the double insertion
\begin{eqnarray}
	\gamma_{k n,l}
	&=&
	- \left[\gamma_{k k'} \delta_{n n'}
		+\delta_{k k'} \gamma_{n n'}\right]
		Z^{-1}_{k' n',l'} \tilde{Z}_{l' l}
	- \left[\mu \frac{\diff}{\diff \mu} Z^{-1}_{kn,l'}\right]
		\tilde{Z}_{l' l}
	.
\label{DefAnomTensor}
\end{eqnarray}
Using the perturbative expansions for the renormalization constants
\begin{eqnarray}
	Z^{-1}_{kn,l}
	&=&
	\sum_{j}
	\left(\frac{g^2}{16\pi^2}\right)^{j}
	Z^{-1,\left(j\right)}_{kn,l},
	\hspace{1cm}
	Z^{-1,\left(j\right)}_{kn,l}
	=
	\sum_{i=0}^{j}
	\frac{1}{\eps^{i}}
	\left[Z^{-1,\left(j\right)}_{i}\right]_{kn,l}
\end{eqnarray}
and $\tilde{Z}$ we derive the perturbative expression for
%$\gamma_{kn,l}$
\begin{eqnarray}
	\gamma_{kn,l}
	&=&
	\frac{g^2}{16 \pi^2} \gamma_{kn,l}^{\left(0\right)}
	+
	\left(\frac{g^2}{16 \pi^2}\right)^{2}
		\gamma_{kn,l}^{\left(1\right)}
	+
	\ldots
\label{AnomDimTens}
\end{eqnarray}
in \eq{DefAnomTensor} up to NLO:
\begin{eqnarray}
	\gamma^{\left(0\right)}_{kn,l}
	&=&
	2 \left[Z^{-1,\left(1\right)}_{1}\right]_{kn,l}
	+ 2 \eps \left[Z^{-1,\left(1\right)}_{0}\right]_{kn,l}
\nn \\
	\gamma^{\left(1\right)}_{kn,l}
	&=&
	4 \left[Z^{-1,\left(2\right)}_{1}\right]_{kn,l}
	+ 2 \beta_{0} \left[Z^{-1,\left(1\right)}_{0}\right]_{kn,l}
\nn \\
	& &
	- 2 \left[Z^{-1,\left(1\right)}_{0}\right]_{kn,l'}
		\left[\tilde{Z}^{-1,\left(1\right)}_{1}\right]_{l' l}
	- 2 \left[Z^{-1,\left(1\right)}_{1}\right]_{kn,l'}
		\left[\tilde{Z}^{-1,\left(1\right)}_{0}\right]_{l' l}
\nn \\
	& &
	- 2 \left\{
		\left[Z^{-1,\left(1\right)}_{0}\right]_{k k'}
		\delta_{n n'}
		+
		\delta_{k k'}
		\left[Z^{-1,\left(1\right)}_{0}\right]_{n n'}
	\right\}
		\left[Z^{-1,\left(1\right)}_{1}\right]_{k' n',l}
\nn \\
	& &
	- 2 \left\{
		\left[Z^{-1,\left(1\right)}_{1}\right]_{k k'}
		\delta_{n n'}
		+
		\delta_{k k'}
		\left[Z^{-1,\left(1\right)}_{1}\right]_{n n'}
	\right\}
		\left[Z^{-1,\left(1\right)}_{0}\right]_{k' n',l}
	.
\label{gamma1double}
\end{eqnarray}
The indices run over both physical and evanescent operators.
The reader may have noticed, that we have used the perturbative
expansions of $Z^{-1}$, $\tilde{Z}^{-1}$ rather than $Z$, $\tilde{Z}$
as in the previous sections.
This is more convenient for the case of double insertions.
Using these equations, the finite renormalization ensuring
\eq{CondMatchDouble} to hold and the locality of counterterms, one
shows in complete analogy to section~\ref{Sect:Triang}:
\begin{eqnarray}
	\gamma^{\left(0\right)}_{E_{rk}l,n}
	&=&
	\gamma^{\left(1\right)}_{E_{rk}l,n}
	=
	0
	\hspace{1cm}
	\mbox{and}
	\hspace{1cm}
	\gamma^{\left(0\right)}_{E_{rk}E_{sl},n}
	=
	\gamma^{\left(1\right)}_{E_{rk}E_{sl},n}
	=
	0
	,
\end{eqnarray}
i.e.\ a double insertion containing at least one evanescent operator
does not mix into physical operators.
Together with the statement that evanescent operators do not
contribute to the matching this proves our method to be consistent at
the NLO level.
As in the case of single insertions one can pick the $\tilde{a}_{kl}$,
$\tilde{b}_{kl}$,\ldots
completely arbitrary and then has to perform a finite
renormalization for the double insertions containing an evanescent
operator in \eq{DoubleWithEva}.
This statement remains valid also in higher orders of the SU(N)
interaction, which can be proven analogously to the proof given
by Dugan and Grinstein \cite{dg} for the case of single insertions.

Now we use the findings above to show the nonvanishing terms
in \eq{gamma1double} explicitly for the physical submatrix:
\begin{eqnarray}
	\gamma^{\left(1\right)}_{kn,l}
	&=&
	4 \left[Z^{-1,\left(2\right)}_{1}\right]_{kn,l}
	- 2 \left[Z^{-1,\left(1\right)}_{1}\right]_{kn,E_{1 l'}}
		\left[\tilde{Z}^{-1,\left(1\right)}_{0}\right]_{E_{1 l'}l}
\nn \\
	& &
	- 2 \left[Z^{-1,\left(1\right)}_{1}\right]_{k E_{1 k'}}
		\left[Z^{-1,\left(1\right)}_{0}\right]_{E_{1 k'} n,l}
	- 2 \left[Z^{-1,\left(1\right)}_{1}\right]_{n E_{1 n'}}
		\left[Z^{-1,\left(1\right)}_{0}\right]_{k E_{1 n'},l}
\nn \\
	& &
\label{physdoub}
\end{eqnarray}
The last equation encodes the following rule for the correct
treatment of evanescent operators in NLO calculations:
{\em The correct contribution of evanescent operators to the NLO
	physical anomalous dimension tensor is obtained by inserting
	the  evanescent one-loop counterterms with a factor of $\frac{1}{2}$
	instead of\/ $1$ into the counterterm graphs.}
Hence the finding of \cite{bw} for a single operator insertion
generalizes to Green's functions with double insertions.
Here the second term in \eq{physdoub} corresponds to the graphs with
the insertion of a local evanescent  counterterm into the graphs
depicted in fig.~\ref{Fig:1}, while the last to terms correspond to
the diagrams of fig.~\ref{Fig:2} with one physical and one evanescent
operator.

\subsection{Double Insertions: Evanescent Scheme Dependences}
\label{Sect:DoubleScheme}

In this section we will answer questions~\ref{Q3} and \ref{Q4} from
page~\pageref{Q3}.
Let us first look at the variation of the anomalous dimension tensor
$\gamma_{kk',l}$ on the coefficients $a_{rs}$.
First one notices, that the LO $\gamma^{\left(0\right)}_{kk',l}$ is
independent of the choice of the $a_{rs}$.
In the NLO case one derives in a way completely analogously to the
procedure presented in section~\ref{Sect:Scheme} the following relation
\begin{eqnarray}
	\gamma^{\left(1\right)}_{kk',l}\left(a'\right)
	&=&
	\gamma^{\left(1\right)}_{kk',l}\left(a\right)
	+
	\left[D \cdot \left(a'-a\right)\right]_{ks}
		\gamma^{\left(0\right)}_{sk',l}
	+
	\left[D \cdot \left(a'-a\right)\right]_{k' s}
		\gamma^{\left(0\right)}_{sk,l}
\label{SchemeAnomBi1}
\end{eqnarray}
with the diagonal matrix $D$ from \eq{DiagMatrix1}.
Note that the indices only run over the physical subspace.

The variation of the anomalous dimension tensor $\gamma_{k k',l}$ with
the coefficients $\tilde{a}_{rs}$ again vanishes in LO, in NLO we find
the transformation
\begin{eqnarray}
	\gamma^{\left(1\right)}_{kk',l}\left(\tilde{a}'\right)
	&=&
	\gamma^{\left(1\right)}_{kk',l}\left(\tilde{a}\right)
	+
	\gamma^{\left(0\right)}_{kk',i}
		\left[\tilde{Z}^{-1,\left(1\right)}_{1}\right]_{i E_{1i}}
		\left[\tilde{a}'-\tilde{a}\right]_{il}
	- 2 \beta_{0}
		\left[Z^{-1,\left(1\right)}_{1}\right]_{kk',\tilde{E}_{1i}}
		\left[\tilde{a}'-\tilde{a}\right]_{il}
\nn \\
	& &
	+ \left[
		\gamma^{\left(0\right)}_{kj}
		\delta_{k' j'}
		+
		\delta_{kj}
		\gamma^{\left(0\right)}_{k' j'}
		\right]
		\left[Z^{-1,\left(1\right)}_{1}\right]_{jj',\tilde{E}_{1i}}
		\left[\tilde{a}'-\tilde{a}\right]_{il}
\nn \\
	& &
	- \left[Z^{-1,\left(1\right)}_{1}\right]_{kk',\tilde{E}_{1i}}
		\left[\tilde{a}'-\tilde{a}\right]_{is}
		\tilde{\gamma}^{\left(0\right)}_{sl}
\end{eqnarray}

As in the case of single insertions, up to the NLO level there exists
no dependence of $\gamma$ on the coefficients $b_{rs}$ and also no one
on the $\tilde{b}_{rs}$.
This provides a nontrivial check of the treatment of evanescent
operators in a practical calculation, when the $b_{rs}$,
$\tilde{b}_{rs}$ are kept arbitrary:
the individual renormalization factors $Z$ each exhibit a dependence
on the coefficients $b_{rs}$, $\tilde{b}_{rs}$ but all this dependence
cancels, when the $Z$'s get combined to $\gamma$.

Next we will elaborate on the scheme independence of RG improved
physical observables.
First look at the solution of the inhomogeneous RG equation \eq{inh}
for the local operator's Wilson coefficient:
\begin{eqnarray}
	\tilde{C}_{l}\!\left(g\left(\mu\right)\right)
	&=&
	\tilde{U}^{\left(0\right)}_{l l'}
		\left(g\!\left(\mu\right),g_{0}\right)
		\tilde{C}_{l'}\!\left(g_{0}\right)
\nn \\
	& &
	+
	\left[\delta_{l l'} +
		\frac{g^{2}\left(\mu\right)}{16\pi^2} \tilde{J}_{l l'}
		\right]
	\cdot
	\int\limits_{g_{0}}^{g\left(\mu\right)} \diff g'\;
	\tilde{U}^{\left(0\right)}_{l' k}
		\left(g\!\left(\mu\right),g'\right)
	\left[\delta_{k k'} -
		\frac{g'^{2}}{16\pi^2} \tilde{J}_{k k'}
		\right]
\nn \\
	& &
	\hspace{1cm}
	\cdot
	\left[\delta_{n n'} +
		\frac{g'^{2}}{16\pi^2} J_{n n'}
		\right]
	U^{\left(0\right)}_{n' t}\left(g',g_{0}\right)
	\left[\delta_{t t'} -
		\frac{g'^{2}}{16\pi^2} J_{t t'}
		\right]
	C_{t'}\!\left(g_{0}\right)
\nn \\
	& &
	\hspace{1cm}
	\cdot
	\left[\delta_{m m'} +
		\frac{g'^{2}}{16\pi^2} J_{m m'}
		\right]
	U^{\left(0\right)}_{m' v}\left(g',g_{0}\right)
	\left[\delta_{v v'} -
		\frac{g'^{2}}{16\pi^2} J_{v v'}
		\right]
	C_{v'}\!\left(g_{0}\right)
\nn \\
	& &
	\hspace{1cm}
	\cdot
	\left\{
		- \frac{\gamma^{\left(0\right)}_{nm,k'}}{\beta_{0}}
			\frac{1}{g'}
		+ \left[
			\frac{\beta_{1}}{\beta_{0}^2}
				\gamma^{\left(0\right)}_{nm,k'}
			-
			\frac{\gamma^{\left(1\right)}_{nm,k'}}{\beta_{0}}
		\right]
		\frac{g'}{16\pi^2}
	\right\}
	.
\label{inhsol}
\end{eqnarray}
Here the matrices $U^{\left(0\right)}$, $\tilde{U}^{\left(0\right)}$
denote the LO evolution matrices stemming from the solution of the
homogeneous RG equations for the Wilson coefficients $C$, $\tilde{C}$,
which reads
\begin{eqnarray}
	U^{\left(0\right)}\left(g,g_{0}\right)
	&=&
	\left[\frac{g_{0}}{g}\right]
		^{ % \displaystyle
			\frac{{\gamma^{\left(0\right)}}^{T}}{\beta_{0}}}
	.
\end{eqnarray}
We have not labeled the evolution matrices with the renormalization
scales $\mu$, $\mu_{0}$ but rather with the corresponding coupling
constants $g\left(\mu\right)$ and $g_{0}=g\left(\mu_{0}\right)$.
The matrix $J$ is a solution of the matrix equation \cite{bjlw}:
\begin{eqnarray}
	J
	+ \left[
		\frac{{\gamma^{\left(0\right)}}^{T}}{2 \beta_{0}},
		J
		\right]
	&=&
	- \frac{{\gamma^{\left(1\right)}}^{T}}{2 \beta_{0}}
	+ \frac{\beta_{1}}{2 \beta_{0}^2}
		{\gamma^{\left(0\right)}}^{T}
	.
\end{eqnarray}
The matrices $\tilde{U}^{\left(0\right)}$, $\tilde{J}$ are defined
analogously in terms of $\tilde{\gamma}$.
If $\gamma$ transforms according to \eq{result}, we know from
\cite{bjlw} that $J$ transforms as
\begin{eqnarray}
	J(a^\prime)
	&=&
	J(a)
	- \left[ D \cdot (a^\prime -a) \right]^{T}
	,
\label{SchemeJ1}
\end{eqnarray}
which can be easily verified from \eq{depma}.
Hence after inserting \eq{SchemeWC1}, \eq{SchemeAnomBi1} and
\eq{SchemeJ1} into \eq{inhsol} one finds the independence of
$\tilde{C}_l$ from the coefficients $a_{kl}$.

In a way similar to the one described above, one treats the scheme
dependence coming from the coefficients $\tilde{a}_{kl}$.
Here some work has been necessary to prove the cancellation of the
scheme dependence connected to $g^{\prime 2} \tilde{J}_{kk^\prime}$ and
$\g^{(1)}_{nm,k^\prime}$ in \eq{inhsol}:
Although it is not possible to perform the integration in \eq{inhsol}
without transforming some of the operators to the diagonal basis, one
can do the integral for the scheme dependent part of \eq{inhsol},
because the part of the integrand depending on $\tilde{a}_{kl}$'s is a
total derivative with respect to $g$.
There is one important difference compared to the case of the
dependence on the $a_{kl}$'s:
A scheme dependence of the Wilson coefficient stemming from the lower
end of the RG evolution remains.
This is a well-known feature of RG improved perturbation theory
\cite{bjlw}.
This residual $\tilde{a}_{kl}$ dependence must be canceled by a
corresponding one in the hadronic matrix element.  If the matrix
elements are obtained in perturbation theory, one can show that the
$\tilde{a}_{kl}$ dependence of the $\tilde{C}_{l}$ gets completely
resolved.  Finally, as in the case of single insertions \cite{bjlw},
one can define a scheme-independent Wilson coefficient for the local
operator
\begin{eqnarray}
	\overline{\tilde{C}_{l}\!\left(\mu\right)}
	&=&
	\left[
		\delta_{l l'}
		+
		\frac{g^2\left(\mu\right)}{16\pi^2}
		\cdot
		\tilde{r}_{l' l}
	\right]
	\tilde{C}_{l'}
	+
	\frac{g^2\left(\mu\right)}{16\pi^2}
	\cdot
	\tilde{r}_{nm,l}
	\cdot
	C^{\left(0\right)}_{n}\!\left(\mu\right)
	C^{\left(0\right)}_{m}\!\left(\mu\right)
	+
	O\left(g^4\right)
	,
\nn \\
	& &
\end{eqnarray}
which multiplies a scheme independent matrix element defined
accordingly.
It contains the analogue of $\hat{r}$ in \cite{bjlw} for the double
insertion
\begin{eqnarray}
	\left\langle \frac{i}{2} \int \T Q_{n} Q_{m}
		\right\rangle^{\left(0\right)}
	&=&
	\frac{g^2}{16\pi^2}
	\cdot
	\tilde{r}_{nm,l}
	\cdot
	\left\langle \tilde{Q}_{l} \right\rangle^{\left(0\right)}
	.
\end{eqnarray}

% here ends SH's work

% here starts the INCLUSIVE section

\section{Inclusive Decays}
\label{Sect:Inclusive}

\begin{figure}[htb]
	\centerline{
	\rotate[r]{
		\epsfysize=5cm
		\epsffile{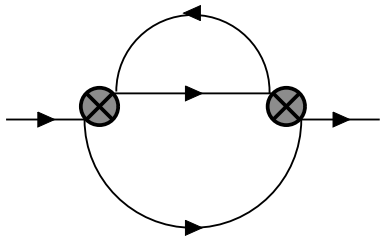}
	}}
	\caption{The lowest order self-energy diagram needed for the
	calculation of inclusive decays via the optical theorem
	(method~2).}
\label{Fig:3}
\end{figure}

Inclusive decays are calculated either by calculating the renormalized
amplitude and performing a subsequent phase space integration and a
summation over final polarizations etc.\ (referenced as method 1)
or by use of the optical theorem, which corresponds to taking the
imaginary part of the self-energy diagram depicted in fig.~\ref{Fig:3}
(method 2).
This figure shows that inclusive decays are in fact related to double
insertions, but in contrast to the case of section~\ref{Sect:Double}
they do not involve local four-quark operators as counterterms for
double insertions.
In fact, even local two-quark operator counterterms would only be
needed to renormalize the real part, but the imaginary part of their
matrix elements clearly vanishes.
The only scheme dependence to be discussed is therefore the one
associated with the $a_{kl}$'s, $b_{kl}$'s, etc., as there are no
$\tilde{a}_{kl}$'s $\tilde{b}_{kl}$'s, etc.\ involved.

To discuss the dependence on the $a_{kl}$'s it is nevertheless
advantageous to consider method~1, i.e.\ the calculation of the
amplitude plus the subsequent phase space integration.
{From} section~\ref{Sect:Scheme} we already know most of the properties
of the RG improved amplitude:
At the upper renormalization scale the properly renormalized
evanescent operators do not contribute and the scheme dependence
cancels.
Further we know the scheme dependence of the (RG improved) Wilson
coefficients at the lower renormalization scale, because with
\eq{SchemeWC1} and \eq{result} we can use the result of \cite{bjlw}.
What we are left with is the calculation of the properly renormalized
operators in perturbation theory, i.e.\ with on-shell external
momenta.
Clearly the form of the external states does not affect the scheme
dependent terms of the matrix elements, they are again given by
\eq{depma} and therefore trivially cancel  between the Wilson
coefficients and the matrix elements, because the scheme dependent
terms are independent of the external momenta.
Since we now have a finite amplitude which is scheme independent, we
may continue the calculation in four dimensions and therefore forget
about the evanescent operators.
The remaining phase space integration and summation over final
polarizations does not introduce any new scheme dependence, therefore
we end up with a rate independent of the $a_{kl}$'s, $b_{kl}$'s, etc.
\footnote{We discard problems due to infrared singularities and the
	Bloch-Nordsiek theorem.
	At least in NLO one can use a gluon mass, because no
	three-gluon vertex contributes to the relevant diagrams}

Alternatively one may use the approach via the optical theorem
(method~2).
Then one has to calculate the imaginary parts of the diagram in
fig.~\ref{Fig:3} plus gluonic corrections.
Of course the properly renormalized operators have to be plugged in:
\begin{eqnarray}
	\imag \left\langle
	\hat{O}^{\ren}_{k} \hat{O}^{\ren}_{l}
	\right\rangle
\end{eqnarray}
One immediately ends up with a finite rate.
What we only have to show is the consistency of the optical theorem
with the presence of evanescent operators and with their arbitrary
definition proposed in \eq{DefEvan1}, \eq{defe2}.
This means that evanescent operators must not contribute to the rate,
i.e.\ diagrams containing an insertion of one or two evanescent
operators must be of order $\eps$
\begin{eqnarray}
	\imag \left\langle
	\hat{E}_{i}\left[O_{k}\right]^{\ren} \hat{O}^{\ren}_{l}
	\right\rangle
	=
	O\left(\eps\right)
	\hspace{0.3cm}
	&\mbox{and}&
	\hspace{0.3cm}
	\imag \left\langle
	\hat{E}_{i}\left[O_{k}\right]^{\ren}
	\hat{E}_{j}\left[O_{l}\right]^{\ren}
	\right\rangle
	=
	O\left(\eps\right)
	.
\label{CondInclNoEva}
\end{eqnarray}
As in the previous sections one can discuss tensor integrals and Dirac
algebra separately leading to \eq{CondInclNoEva}.

% here ends the INCLUSIVE section

\section{Conclusions}
\label{Sect:Concl}
In this work we have analyzed the effect of different definitions of
evanescent operators.  We have shown that one may arbitrarily
redefine any evanescent operator by $(D-4)$ times any physical
operator without affecting the block-triangular form of the
anomalous dimension matrix, which ensures that
properly renormalized evanescent operators
do not mix into physical ones. Especially one is not forced to
use the definition of the evanescent operators proposed in \cite{dg},
whose implementation is quite cumbersome.
Then we have analyzed the
renormalization scheme dependence associated with the
redefinition transformation in the next-to-leading order in
renormalization group improved perturbation theory.
We stress that it is meaningless to give some anomalous dimension
matrix or some Wilson coefficients beyond leading logarithms without
specifying the definition of the evanescent operators used during the
calculation.
In physical observables, however, this renormalization scheme dependence
cancels between Wilson coefficients and the anomalous dimension matrix.
One may take advantage of this feature by defining the
evanescent operators such as to achieve a simple form for the
anomalous dimension matrix.
Then we have extended the work of \cite{bw} and \cite{dg}
to the case of Green's functions with two operator insertions
and have also analyzed the abovementioned renormalization scheme
dependence.
For this we have set up the NLO renormalization group formalism for
four-quark Green's functions with two operator insertions, we have
derived the renormalization scheme dependence of the corresponding
anomalous dimension tensors and defined scheme-independent Wilson
coefficients.
Finally we have analyzed inclusive decay rates.

\section*{Acknowledgements}
The authors thank Andrzej Buras and Miko{\l}aj Misiak for useful
discussions.


\begin{thebibliography}{99}
\bibitem{b} G.~Bonneau,
	Nucl.~Phys.\ {\bf B167}(1980)261,
        Nucl.~Phys.\ {\bf B171}(1980)477.
\bibitem{c} J.~Collins,
	{\sl Renormalization},
	Cambridge Univ.\ Press 1984.
\bibitem{bm} P.~Breitenlohner and D.~Maison,
	Comm.~Math.~Phys.\ {\bf 52}(1977)11.
\bibitem{zim} W.~Zimmermann,
	Ann.~Phys.\ {\bf 77}(1973)536.
\bibitem{bw} A.~J.~Buras and P.~H.~Weisz,
	Nucl.~Phys.\ {\bf B333}(1990)66.
\bibitem{dg} M.~J.~Dugan and B.~Grinstein,
	Phys.~Lett.\ {\bf B256}(1991)239.
\bibitem{bos} M.~Bos,
	Ann.~Phys.\ {\bf 181}(1988)177; \\
        C.~Schubert,
	Nucl.~Phys.\ {\bf 323}(1989)478.
\bibitem{ho} G.~'t~Hooft and M.~Veltman,
	Nucl.~Phys.\ {\bf B44}(1972)189.
\bibitem{m} A.~J.~Buras, M.~E.~Lautenbacher, M.~Misiak, M.~M\"unz,
	Nucl.~Phys.\ {\bf B423}(1994)349.
\bibitem{bjlw} A.~J.~Buras, M.~Jamin, M.~E.~Lautenbacher, P.~H.~Weisz,
	Nucl.~Phys.\ {\bf B370}(1992)69.
\bibitem{hn} S.~Herrlich and U.~Nierste,
	Nucl.~Phys.\ {\bf B419}(1994)292; \\
	S.~Herrlich,
	{\sl QCD Korrekturen h{\"o}herer Ordnung zur
		${\rm K^{0}}$--$\overline{\rm K^{0}}$
		Mischung},
	{\it in German},
	PhD Thesis, Technical University of Munich, 1994.
\bibitem{hn2} S.~Herrlich and U.~Nierste,
	{\sl The complete $\Delta S=2$--hamiltonian in the next-to-leading
		order}, TUM-T31-86/95 ({\it in preparation}); \\
	U.~Nierste,
	{\sl Indirect CP Violation in the Neutral Kaon System Beyond
		Leading Logarithms},
	PhD Thesis, Technical University of Munich, 1995.
\bibitem{bb} G.~Buchalla and A.~J.~Buras,
	Nucl.~Phys.\ {\bf B412}(1994)106.
\end{thebibliography}
\end{document}